\documentstyle[multicol,aps,prl,epsf,psfig]{revtex}
\begin{document}
\title{A possible experimental test of the thermodynamic approach to
granular media}
\author{David S. Dean
and Alexandre Lef\`evre\\ {IRSAMC, Laboratoire de Physique Quantique,
Universit\'e Paul Sabatier, 118 route de Narbonne, 31062 Toulouse
Cedex 04, France} }  \maketitle

\begin{abstract}
We study the steady state distribution of the energy of the
Sherrington-Kirkpatrick model driven by a tapping mechanism which
mimics the mechanically driven dynamics of granular media.  The
dynamics consists of two phases: a zero temperature relaxation phase
which leads the system to a metastable state, then a tapping which
excites the system thus reactivating the relaxational dynamics.
Numerically we investigate whether the distribution of the energies of
the blocked states obtained agrees with a simple canonical form of the
Edwards measure. It is found that this canonical measure is in good
agreement with the dynamically measured energy distribution. A possible
experimental test of the Edwards measure based on the study here is proposed.
\vskip 0.5cm
\date{November 25 2002}

\end{abstract}

\begin{multicols}{2}
Complex systems such as granular media possess a large number of
metastable or blocked configurations. When a granular medium is shaken
it quickly relaxes into a blocked configuration, a subsequent shake or
tap will lead it to another blocked or jammed state and so on. If the
driving mechanism is held constant one expects the system to enter
into a quasi-equilibrium stationary state.  Various driving mechanisms
can be investigated experimentally, such as vertical tapping
\cite{vert} and horizontal shaking \cite{horiz}.  In granular media
and other complex systems such as spin glasses, the entropy of these
blocked states is extensive in the system size, and hence it has been
proposed that one may use a thermodynamic measure over blocked states
to describe this steady state. The simplest proposition is that the
system is characterized by a number of quantities which are fixed on
average and then the measure on the steady state is obtained from the
maximum entropy state (on blocked states) with the relevant
macroscopic quantities fixed \cite{edmes}. This simple idea has
recently been investigated in a wide range of systems and has been
shown to be relatively successful.Various tests of the applicability
of these thermodynamic ideas have been carried out and although some
confirmation has been made in more realistic sheared granular systems
\cite{maku}, most work has been carried out on simpler model systems
which one hopes capture the basic physics of granular media.  The
Edwards flat measure has been shown to be of predictive value in some
simple one dimensional lattice models \cite{brey,oned1}, but there are
clearly examples where the approach fails \cite{bfs}. However, recently
it was shown that more sophisticated versions of the Edwards measure
introducing ensembles with several quantities fixed on average can
remedy the deficiencies of the basic measure in these cases
\cite{alex}.  Other toy models that have been analyzed are lattice
based models with kinetic constraints in higher dimensions
\cite{bar,lattice} and also spin glass models \cite{sgs} where non-thermal
driving is used to move the system between blocked states.
 
Even if it is not expected to be exact, many systems may be described
to a good engineering level by these measures. Given the difficulty of
the analysis of the highly nonlocal dynamics in these systems this is
an important step toward understanding their steady state
regimes. There is no clear ergodicity in these systems and no detailed
balance as in usual statistical mechanics. Edwards argued that a
system might conceivably explore blocked configurations in a flat
manner if the driving involved extensive manipulations, meaning the
displacement of a macroscopic number of particles, for example
shaking, stirring or pouring granular media. An interesting
consequence of the applicability of thermodynamic ideas is that one
may describe phase transitions in these driven systems \cite{trans}.
However, even considering the success of the Edwards measure in describing
various simple models, evidence in realistic granular media is still
lacking. In this Letter, we investigate whether, at fixed tapping rate (to be
defined later), the states explored dynamically obey a form of Boltzmann
distribution. The results presented here are quite striking, despite a lack
of detailed balance we shall see that a Boltzmann distribution excellently
describes the histogram of the energies of the blocked states visited
during the tapping. Motivated by these results, we propose a generic and 
simple experimental test of the Edwards measure, which should be feasible in
a wide range of driven granular systems.

As mentioned above, a good theoretical and numerical testing ground for
this thermodynamic approach to granular media are spin glasses. Spin
glasses have an exponentially large (with the system size) number of
metastable states and hence an extensive entropy of blocked states as
do granular media. The definition of a blocked state in a spin glass
simulated on a computer depends of course on the local dynamics. Under
single spin flip dynamics a metastable state is one where flipping any
single spin increases the energy, it is thus a blocked state under any
single spin flip Monte Carlo dynamics. Various spin glass models have
been studied to explore the accuracy of the Edwards measure as a
function of the relaxational dynamics and the tapping mechanism
\cite{sgs}.  Here we shall explore the driven dynamics of the
Sherrington-Kirkpatrick (SK) \cite{sk} spin glass model defined by the
Hamiltonian $H = -\sum_{i<j} J_{ij} S_i S_j$,
where the $S_i$ ($1\leq i\leq N$) are Ising spins and the interaction
$J_{ij}$ are independent Gaussian random variables of zero mean and
variance $\overline{J_{ij}^2} = 1/N$.

The tapping mechanism introduced in spin glass models involves two
steps: 1) Relaxation: the system evolves under zero temperature Glauber
dynamics until it is blocked in a metastable state; 2) The system is
tapped with tapping parameter $p$, that is to say each spin is flipped
in parallel with probability $p$ - this is the extensive manipulation
as at each tap on average $pN$ spins are flipped.  Given a 
tapping rate $p$, it is observed that the system reaches a steady state
regime with a fixed average energy per spin $\epsilon$ (the energy being
measured at the end of step 1) above). If we
postulate the canonical form of the Edwards measure for the
quasi-equilibrium distribution of the driven dynamics (observed on the
blocked states) one has the steady state distribution of energy given
by
\begin{equation}
\rho_{EDW}(E) = {N_{MS}(E) \exp(-\beta E) \over Z}\label{test}
\end{equation}
where $N_{MS}(E)$ is the number of metastable states of energy $E$,
$\beta$ is the inverse Edwards temperature, which can be thought of as
a Lagrange multiplier fixing the average energy per spin, and $Z=\sum_{E} N_{MS}(E) \exp(-\beta
E)$ is the canonical partition function. We note that a fully flat measure 
with fixed energy per spin
would be $P_\alpha = \exp(-\beta E_\alpha)/Z$, where $E_\alpha$ is the
energy of the blocked state $\alpha$. In this paper we shall
investigate only the weaker form of the measure Eq. (\ref{test}). Such an
investigation has been carried out in a model of particle deposition \cite{brey}, 
but the originality of the approach of this Letter is that 
one can partially
test Eq. (\ref{test}) without knowing the Edwards entropy. This is crucial,
since it opens up the possibility of looking at a wider range of granular
models and even experiments.

The hypothesis Eq. (\ref{test}) will be tested in the following
manner. We fix a bin size for the energies $\Delta E$ and by exact
enumeration we compute $N_{MS}(E)$ for a given system and realization
of the disorder. The exact enumeration means that we can explore
system sizes up to $N=30$. Then the tapping dynamics is applied to the
same system and the dynamical histogram $N_{D}(E,p)$ of the energy
with the same bin size $\Delta E$ is computed over $10^6$ taps after
the system has been tapped $10^6$ times to reach the steady state.
If the Edwards measure in the form Eq. (\ref{test}) is
valid then one should obtain
\begin{equation}
r(E)=\ln({N_{D}(E,p)\over N_{MS}(E)}) = -\beta E + C \label{log}
\end{equation} 
where $C$ is a constant which should be independent of $E$. 
If Eq. (\ref{log}) is valid then an interesting question to
ask is what is the dependence of $\beta$ on a) the tapping strength
$p$, b) the size of the system $N$ c) the realization of the disorder.
It is interesting to see if the tapping strength $p$ determines a
unique temperature in the limit of large system sizes, for systems
with the same statistical distribution of couplings {\em i.e.}  is
$\beta$ self averaging? We shall see that the results obtained here
for small system sizes suggest that Eq. (\ref{log}) is obeyed on
increasing the system size and that the parameter $\beta$ becomes
independent of the system size and the disorder realization for large
systems.

In order to see what happens for larger systems, where the enumeration
of metastable states is no longer feasible and where the thermodynamic
approach should be expected to work better, one may compare the
dynamical distribution $N_D(E,p)$ measured for the steady states at
different tapping rates. If the Edwards measure holds, {\em i.e.}
$N_D(E,p)\propto \rho_{EDW}(E)$, then for any two tapping strengths $p$
and $p'$ we should find that
\begin{equation}
{N_D(E,p)\over N_D(E,p')} = \exp\left(-(\beta(p)-\beta(p'))E+f(N,p,p')\right)
\end{equation}
or equivalently
\begin{eqnarray}
c(p,p',\epsilon)&=&{\ln(N_D(E,p)) - \ln(N_D(E,p'))\over N} \nonumber
\\ &=& -(\beta(p)-\beta(p'))\epsilon+\frac{1}{N} f(N,p,p') \label{eqcpp}
\end{eqnarray}
where $\epsilon = E/N$ is the energy per spin and $f(N,p,p')$ is independent of
$E$. This constitutes
a weak test of the Edwards hypothesis but has the merit of
being practicable for large system sizes.

A final test is to use the analytic, but annealed calculation, of the
average number of metastable states at fixed energy,
$\overline{N_{MS}(\epsilon)}$ (the overline indicates the disorder
average over realizations), which gives in the thermodynamic limit
$\overline{N_{MS}(\epsilon)} \approx \exp(N S^a_{MS}(\epsilon))$, where
\cite{taed,bm} 
\begin{equation}
S^a_{MS}(\epsilon) = \min_{z}\left\{{1\over 2} z^2 - E^2 + \ln\left( 1
- {\rm erf}\left({z + 2 E \over \sqrt{2}}\right)\right)\right\}
\end{equation}
is the annealed entropy per spin of metastable states of energy
$\epsilon$ per spin. If the Edwards measure holds then we should find that
\begin{equation}
{\ln\left(N_D(\epsilon,p)\right)-\ln\left(N_{MS}(\epsilon)\right)
\over N} = -\beta(p)\epsilon + C
\label{largen}
\end{equation}
We further expect that the dynamical entropy per spin
$S_D(\epsilon,p)=\ln(N_D(\epsilon,p))/N$ and the entropy of metastable
states per spin $S_{MS}(\epsilon)=\ln(N_{MS}(\epsilon,p))/N$ are self
averaging and thus Eq. (\ref{largen}) may be written for large $N$ as
$r(\epsilon) = \overline{S_D(\epsilon,p)} - \overline{S_{MS}(\epsilon)} =
-\beta(p)\epsilon + C$.
Unfortunately no exact calculation for $\overline{S_{MS}(\epsilon)}$
exists, however it is known that $S^a_{MS}(\epsilon) =
\overline{S_{MS}(\epsilon)}$ for $\epsilon_c >-0.672$
\cite{bm}. Furthermore this annealed approximation seems to be good,
though not exact, to even lower energies as confirmed by a replica
symmetric calculation \cite{rob}. Hence at energies above $
\epsilon_c$ we have the formula
\begin{equation}
r_a(\epsilon)=
\overline{S_D(\epsilon,p)} -{S^a_{MS}(\epsilon)} = -\beta(p)\epsilon + C
\end{equation}
which should also work well approximately at energies below, but not
too far from, $\epsilon_c$. We remark here that by Jensen's inequality
$S^a_{MS}(\epsilon) \geq S_{MS}(\epsilon)$.

\begin{figure}
\narrowtext \epsfxsize=1.0\hsize \epsfbox{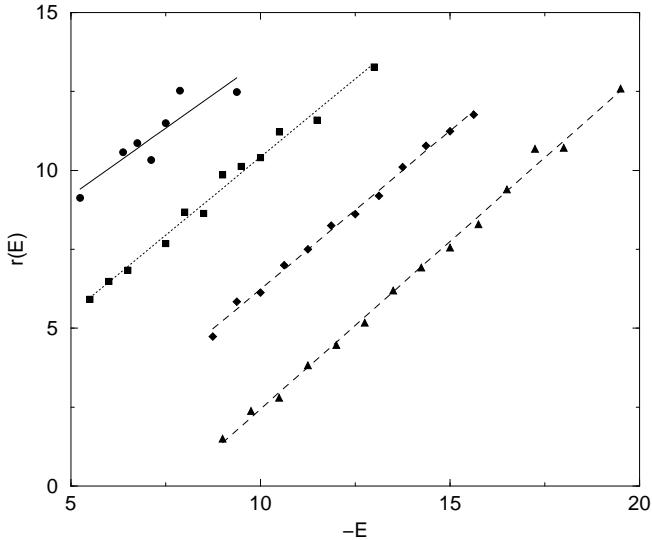}
\caption{The computed values of $r(E)$ for the randomized spin flip
relaxation at $p=0.3$ with a bin size of $0.025 N$ for systems sizes
$N$ of 15 (circles), 20 (squares), 25 (diamonds) and 30
(triangles). The straight lines are linear fits to guide the eye.}
\label{fig1}
\end{figure}
We now discuss the results of these numerical tests. Systems between 
size $15$ and $30$ were analyzed. The numerically measured value of
$r(E)$ was computed from the static metastable state energy histogram
and the histogram obtained by tapping the system $10^6$ times. This
was done for values of $p$, 0.1, 0.2, 0.3, 0.4 and 0.5. Increasing the 
number of taps did not change the dynamical histogram so we can 
be sure that we are measuring the steady state (this is natural given the 
small system sizes). It is possible that with very low tapping rates
the time to reach the steady state even for a system of $30$ spins becomes 
large. As an example the results for $p=0.3$ are shown in Fig. (\ref{fig1}).
We see
that for small systems sizes the points of the numerically measured $r(E)$ are 
scattered about the straight line fit predicted by Eq. (\ref{log}). However
as the system size is increased the straight line fit appears excellent. The
deviations from the straight line fit also appear to be non systematic. 
Similarly good fits are also obtained for all the tapping rates
studied. In the case of the SK model, a Boltzmann distribution of 
the energies of the metastable states visited in the steady state
regime appears to depend simply on having a large enough system size.
It is generally believed that the Edwards measure should be more accurate in 
gently driven systems, here the weak form characterized by Eq. (\ref{log})
seems equally valid at all tapping rates. We also note that the slope of the
straight lines appear to be saturating to a constant value of $\beta(p)$
on increasing $N$,
suggesting that in the thermodynamic limit the 
Edwards temperature is fully characterized by the tapping rate $p$.

From this first numerical study one may conclude that if the weak form
of the Edwards measure holds then it relies on the system being
sufficiently large. The results above however seem encouraging, especially
when one takes into account that the average total number of metastable states
in a system of size $N$ $\sim$ $\exp(0.1992 N)$ \cite{taed}, 
which is only about $400$ for a system of size $30$.

\begin{figure}
\narrowtext \epsfxsize=1.0\hsize \epsfbox{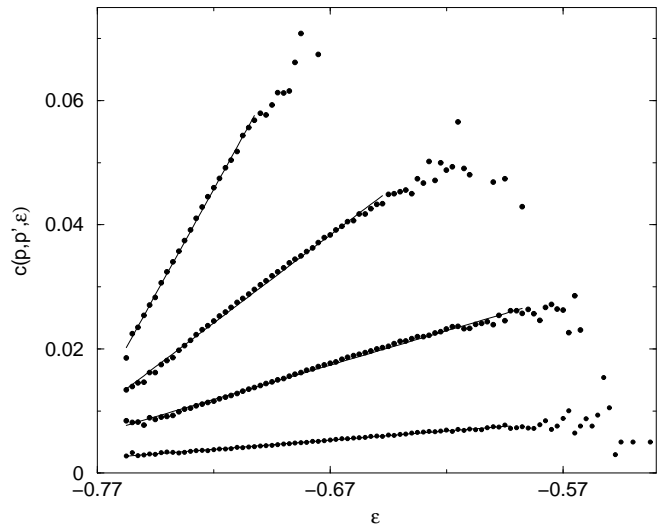}
\caption{The numerically computed values of $c(p,p'\epsilon)$ from
simulations of a system of size $N=200$ spins. The curves have been vertically
shifted for clarity. Shown in descending order are the results for
$(p,p')$: (0.2,0.1), (0.3,0.2), (0.4,0.3), (0.5,0.4).}
\label{fig2}
\end{figure}

We now carry out simulations for a system of size $200$, here the
average number of metastable states is about $2\times 10^{17}$.
The bin size is also refined to $0.0025N$.
Naively one would expect the thermodynamics approach to work better
in this limit as the phase space of metastable states is much bigger.
Also there are many more states in each energy bin and hence 
it is really the weak form of the Edwards measure that is tested.
Shown on Fig. (\ref{fig2}) is the numerically computed $c(p,p,'\epsilon)$
from a simulation of the same system tapped at values of $p$, 0.1, 0.2, 0.3,
0.4 and 0.5. The system was first equilibrated by tapping $5000$ times
and the dynamical histogram was constructed over $5\times 10^6$ taps.
One sees from Fig. (2) that the values of $c(p,p',\epsilon)$ are
to a good degree of accuracy on a straight line, as predicted
by Eq. (\ref{eqcpp}). The deviations at low energies, where the 
sampling is lowest, appear non-systematic.          

\begin{figure}
\narrowtext \epsfxsize=1.0\hsize \epsfbox{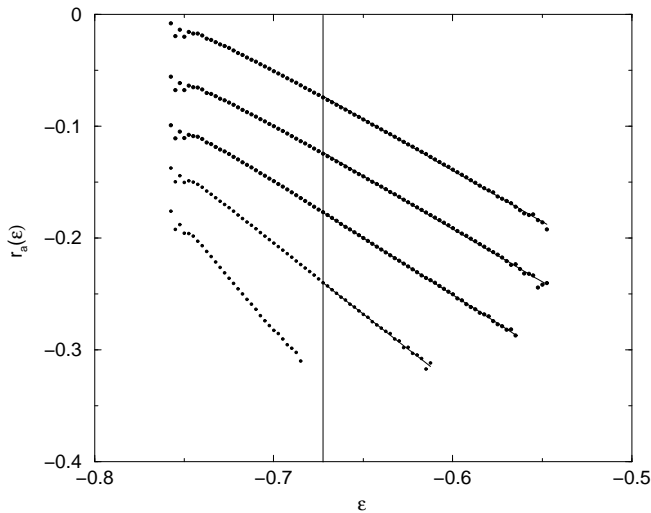}
\caption{The calculation of $r_a(\epsilon)$ for a system with $N=200$
and data taken over $5\times 10^6$ taps. Shown in ascending order are the
plots for $p$: 0.1, 0.2, 0.3, 0.4 and 0.5. The straight line fits are
shown in the region $\epsilon > \epsilon_c$ 
(to the right of the vertical line).}
\label{fig3}
\end{figure}

The energy histograms generated by the precedent simulations were
used to compute $r_a(\epsilon)$, the results are shown in Fig. (\ref{fig3}). 
The curves have been shifted vertically for clarity.
A straight line fit was performed in the region $\epsilon >\epsilon_c$
and we see that in this region the fit is excellent. In addition the values
obtained from the fits for $\beta(p)$ are compatible with those obtained for
the systems of size $N=30$. The lower energy part
of the curves also appear linear down to around $\epsilon = -0.72$.
The replica symmetric
calculation of \cite{rob} also visibly (on the same scale) departs
from the annealed calculation near this energy, although we emphasize
that this is not an exact calculation but can be expected to be closer to
the real result than the annealed one. 
The lower tapping rates explore the lower
energy regime while the higher rates explore the higher energy regime.
The deviations above $\epsilon_c$ from the annealed entropy are at the
high energy end of the numerical histogram where the sampling is smallest.

To summarize we see that the assumption of an effective Boltzmann distribution
for the energy of the states explored by tapping dynamics describes extremely
well the numerically obtained results. This description seems to work better
on increasing the system size for small systems and seems to be valid for 
larger systems via slightly more indirect tests. Why it works is rather
mysterious and if it is exact it would be an extremely useful method
to numerically map out the distribution of metastable states 
in various spin glass models. The relaxational dynamics used was random
update which uses a lot of computer time for these simulations. Sequential
update is much quicker for these simulations and we have carried out
such simulations and found to within very small deviations the same results.
If one could prove the applicability of the Edwards measure for these
types of dynamics one would have an extremely powerful method to explore
metastable states and inherent states in glassy systems. The standard
method employing an auxiliary Hamiltonian \cite{bar}
is much more time consuming - although classical statistical 
mechanics tells us that it will give the right result. 

In the original proposition of Edwards in the context of granular
media, the steady state volume fraction $V$ occupied 
by a driven granular media is described in the canonical approach by
\begin{equation}
\rho_{DYN}(V) \propto \exp(-{V\over X} + S(V))
\end{equation}
where $X$ is the compactivity and $S(V)$ is the entropy of blocked
states occupying volume $V$. The problem of calculating $S(V)$ in a
real system seems formidable. However it should be possible to test
the prediction of Eq. (\ref{eqcpp}) experimentally by comparing
the histograms of $V$ for systems driven at different driving amplitudes.
This would be a crucial test, though we emphasize again, not a demonstration
of the validity of the Edwards measure.

\end{multicols}

\end{document}